\documentclass[aps,pra,reprint,floatfix,showpacs,twocolumn]{revtex4-1}
\usepackage[utf8]{inputenc}
\usepackage{amsfonts}
\usepackage{amssymb}
\usepackage{amsmath}
\usepackage{amstext}
\usepackage{graphicx}
\usepackage{wasysym}
\newcommand{\ie}{\textit{i.e.}~}
\newcommand{\eg}{\textit{e.g.}~}

\renewcommand{\i}{{\rm i}}
\newcommand{\e}{{\rm e}}
\renewcommand{\d}{{\rm d}}

\renewcommand{\o}[2][]{\hat{#2} ^{\vphantom{\dagger}}_{#1}}
\newcommand{\op}[2][]{\hat{#2} ^{\dagger}_{#1}}
\newcommand{\oo}[2][]{\hat{#2} ^{\vphantom{\dagger}2}_{#1}}
\newcommand{\oop}[2][]{\hat{#2} ^{\dagger 2}_{#1}}

\begin{document}
\title{Second Josephson excitations beyond mean field as a toy model for thermal pressure: exact quantum dynamics and the quantum phase model}

\author{M.\,P.\,Strzys}
\email{strzys@physik.uni-kl.de}
\author{J.\,R.\,Anglin}
\affiliation{OPTIMAS Research Center and Fachbereich Physik, Technische Universit\"at Kaiserslautern, D--67653 Kaiserslautern, Germany}


\begin{abstract}
A simple four-mode Bose-Hubbard model with intrinsic time scale separation can be considered as a paradigm for mesoscopic quantum systems in thermal contact. In our previous work we showed that in addition to coherent particle exchange, a novel slow collective excitation can be identified by a series of Holstein-Primakoff transformations. This resonant energy exchange mode is not predicted by linear Bogoliubov theory, and its frequency is sensitive to interactions among Bogoliubov quasi-particles; it may be referred to as a second Josephson oscillation, in analogy to the second sound mode of liquid Helium II. In this paper we will explore this system beyond the Gross-Pitaevskii mean field regime. We directly compare the classical mean field dynamics to the exact full quantum many-particle dynamics and show good agreement over a large range of the system parameters. The second Josephson frequency becomes imaginary for stronger interactions, however, indicating dynamical instability of the symmetric state. By means of a generalized quantum phase model for the full four-mode system, we then show that, in this regime, high-energy Bogoliubov quasiparticles tend to accumulate in one pair of sites, while the actual particles preferentially occupy the opposite pair. We interpret this as a simple model for thermal pressure.
\end{abstract}

\maketitle

\section{Introduction}
Thermodynamics remains one of the most complete and powerful physical theories in our possession. The connection to statistical physics established in the 19$^{\rm th}$ century was a very effective first attempt to understand the microscopical foundations of thermodynamics and to combine this highly effective theory with classical mechanics. Yet even today the exact relationship between thermodynamics and microphysics is not completely understood. Contemporary experiments investigating small quantum systems, for example Bose-Einstein condensates in optical lattices, are a recent advance that revives old questions. The process of equilibration and thermalization in such small and isolated systems has received considerable theoretical re-examination \cite{Berm05,Berm04,Gira60,Kino06,Rigo07}, and motivated the proposal of new concepts of equilibrium \cite{Deut91,Sred94,Rigo08,Rigo09a,Pono11a,Pono11b}. 

The approach we wish to pursue in this context is based on the view that thermodynamics is ultimately an adiabatic theory, in the dynamical sense of deriving effective Hamiltonians by eliminating high-frequency degrees of freedom. This adiabatic basis of the subject is often concealed, in statistical mechanics, by the definition of ensembles; but the fundamental justification of ensembles lies only in representing time-averaging. For small systems, however, it may become feasible to return to the Hamiltonian and apply adiabatic analysis directly. Such an approach may then shed light on the actual microphysics underlying  thermodynamics, and on its basic relationship with quantum mechanics. We therefore begin this ambitious program for quantum thermodynamics with its simplest possible example, and consider a minimal model Hamiltonian for two small quantum systems in thermal contact.

At least if we restrict attention to models that somewhat resemble the confined gases to which classical thermodynamics traditionally refers, the smallest nontrivial quantum system is surely the two-mode Bose-Hubbard (BH) model. This simple model has been extensively studied, not only theoretically \cite{Smer97,Milb97,Ragh99,Vard01b}, but also experimentally, since it is accessible in many cold atom laboratories \cite{Bloc05,Gati07,Myat97,Matt99}. Weakly coupling two of these simple systems together might therefore be the simplest case of two quantum systems in contact, such that they can exchange both energy and particles independently. In our previous work \cite{Strz10} we introduced a four-mode BH system with a clear time scale separation as such an example of two exactly similar quantum subsystems in thermal contact, and showed how something similar to heat exchange may be described in familiar quantum mechanical terms. 

Each subsystem in this model is itself two-mode BH model -- a pair of bosonic modes coupled so as to represent an idealized Josephson junction (JJ) \cite{Milb97,Ragh99,Giov00,Levy07,Truj09}. We specify that the subsystems' characteristic timescale--their Josephson frequency--is to be the shortest one in the entire system, so that observation and control can address only much longer time scales. This leads to a Hamiltonian of the form
\begin{align}\label{Hamiltonian}
 \o{H} &= \o[\rm L]{H}+\o[\rm R]{H} + \o[\rm LR]{H}\nonumber\\
\o[\rm LR]{H}  &=-\frac{\omega}{2}\left(\op[1{\rm L}]{a}\o[1{\rm R}]{a} + \op[2{\rm L}]{a}\o[2{\rm R}]{a} + \textrm{H. c.}\right),\nonumber\\
\o[\alpha]{H} &=  -\frac{\Omega}{2}\left(\op[1\alpha]{a}\o[2\alpha]{a} +\op[2\alpha]{a}\o[1\alpha]{a} \right) + U\sum_{i=1}^{2} \oop[i\alpha]{a}\oo[i\alpha]{a},
\end{align}
with on-site interaction $U$ and index $\alpha={\rm L,R}$ for the left and right subsystem, respectively. Coupling between the two subsystems is modeled by an additional Josephson tunneling between them with a much smaller rate coefficient $\omega$ than that of the intra-subsystem tunneling $\Omega$ to implement time scale separation ($\omega / \Omega \ll 1$). 

In \cite{Strz10} we showed that although only the total particle number $N$ is an exactly conserved quantum number, in adiabatic theory there exists another conserved quantity, the total number of Bogoliubov exciatations $J$ of both of the subsystems. We called these excitations `josons', since they are the elementary excitations of Josephson junctions. We showed that although linear Bogoliubov theory identifies only one low-frequency normal mode in this system, in fact there exists a second low-frequency collective mode in this adiabatic limit, which emerges from the beating of the two high-frequency Bogoliubov modes, and corresponds to the exchange of josons between the subsystems. Thus, the two subsystems can exchange particles by ordinary (`first') Josephson oscillations on a `slow' time scale, but they can also exchange josons (\ie heat, in the sense of high-frequency energy), as the amplitudes of their fast Josephson oscillations beat slowly back and forth. We call this second low-frequency collective mode `second Josephson oscillations', by analogy with first and second sound in liquid Helium II. 

While the second Josephson mode is not an elementary excitation within the full theory, since it does not exist as a perturbation of the zero-joson ground state, it does appear as an elementary excitation within the adiabatic effective theory that describes low-frequency excitations around highly excited stationary states. Moreover, post-Bogoliubov nonlinear effects that are resummed in the adiabatic analysis shift the second Josephson frequency away from the simple Bogoliubov beat frequency, giving it a nontrivial collective character. Within the low-frequency effective theory, however, the two Josephson modes are described in exactly the same way, with a low-frequency Hilbert space spanned by quantized excitations of both.Thus, our combined four-mode BH system may be regarded as a simple toy model for quantum heat exchange. 

Our analysis of these second Josephson oscillations included the derivation of an extended free quasiparticle theory for collective excitations of the system considered, simply by generalizing the Bogoliubov quasiparticle to treat adiabatically invariant quasiparticle numbers as fixed. Despite the fact that our derivations were made in terms of operators, the whole procedure was in principle classical, since we restricted ourselves to the regime of large $N$ and the limit $J\ll N$. Thus our results in \cite{Strz10} could be presented using nonlinear Gross-Pitaevskii mean field simulations.

In the present paper we now adopt two different approaches to go beyond this mean field regime. In the first place we extend our numerical analysis to the exact full quantum dynamics of the four-mode system. The total number of particles $N$ is restricted by computational effort, but it is achievable to solve cases with $N \gtrsim 40$. While the large $N$ theory of \cite{Strz10} requires corrections for small $N$ or for $J\sim N$ (presented elsewhere \cite{Strz12}), we show here that the quantum results do indeed agree with the mean field approximations for larger $N$.  

Secondly, then, we extend the standard quantum phase model (QPM) for Josephson junctions \cite{Scho90,Legg91,Java99b,Smer00,Pita03} to the two coupled junctions of the four-mode system. This can easily be achieved by requantizing the classical canonical conjugate variables. This model, being also only valid for large $N$ (although it might be generalized for all values of $N$ in the spirit of \cite{Angl01b}), nevertheless is able to predict small quantum corrections to the mean field solutions, such as quantum depletion. The QPM can also conveniently describe so-called `running' states, in which josons preferentially are in one half of the system. We will show that in this regime the concentrated josons effectively repel the atoms, pushing a proportion of them into the other half of the system. We interpret this effect thermodynamically in terms of josons providing pressure, similar to the thermal pressure of a macroscopic ideal gas. 


\section{Review of the joson theory for the four-mode system}\label{section:joson}

As a first step we implement a number-conserving Bogoliubov theory for the two subsystems. To that end we define atom-moving operators $\op[\alpha]{a},\o[\alpha]{a}$ by a Holstein-Primakoff transformation (HPT) \cite{Hols40} of the form
\begin{align}\label{schwing}
	2\sqrt{\o[\alpha]{N}-\op[\alpha]{a}\o[\alpha]{a}}\ \o[\alpha]{a} &\equiv (\op[1\alpha]{a}+\op[2\alpha]{a})(\o[1\alpha]{a}-\o[2\alpha]{a}).
\end{align}
These new operators commute with $\o[\alpha]{N}=\sum_{i=1,2}\op[i\alpha]{a}\o[i\alpha]{a}$ and fulfill the usual bosonic commutation relations $[\o[\alpha]{a},\op[\alpha]{a}]=1$. Thereupon we 
perform a Bogoliubov transformation by letting $\o[\alpha]{a} = u_{\alpha}\o[\alpha]{b} + v_{\alpha}\op[\alpha]{b}$ and choosing $u_{\alpha},v_{\alpha}$ so as to diagonalize the quadratic terms of $\o[\alpha]{H}$ in \eqref{Hamiltonian} while demanding $[\o[\alpha]{b},\op[\alpha]{b}]=1$. Being interested in dynamics slow compared to the timescale set by $\Omega$, \ie an adiabatic regime, we now apply a rotating wave approximation (RWA) and ignore all terms that do not commute with the leading term proportional to $\op[\alpha]{b}\o[\alpha]{b}$ in $\o[\alpha]{H}$. In these terms the subsystems' Hamiltonian reads
\begin{align}\label{Halpha1}
\o[\alpha]{H} = &-\frac{\Omega}{2}\o[\alpha]{N}+\frac{U}{2}\o[\alpha]{N}(\o[\alpha]{N}-2) + \sqrt{\Omega(\Omega+2U\o[\alpha]{N})}\ \op[\alpha]{b}\o[\alpha]{b} \nonumber\\
&-\frac{U}{4}\frac{4\Omega +2U\o[\alpha]{N}}{\Omega + 2U \o[\alpha]{N}}\oop[\alpha]{b}\oo[\alpha]{b} + \mathcal{O}(UN_\alpha^{-1})\;.
\end{align}
Note that the number-conserving Bogoliubov formalism introduced here via an HPT is somewhat different from the  standard number-conserving Bogoliubov theory \cite{Cast98,Gard97,Gard07,Oles08}, but follows quite naturally from the properties of the HPT.

In the new variables the Hamiltonian of the subsystems \eqref{Halpha1} is diagonal. The coefficient of the quadratic term is the well known Josephson frequency of the excitations created by $\op[\alpha]{b}$, which we call `josons', since the are the elementary excitations of the two Josephson junctions. The coefficient of the term quartic in $\o[\alpha]{b}$ is in general of the same order as $U$, but has opposite sign. This implies the simple but somewhat surprising fact that when atoms repel each other, josons attract each other; and \textit{vice versa}.

The total Hamiltonian of the whole four-mode system in zeroth order in $\omega/\Omega$ is just equal to $\tilde\Omega\hat{J}$, where $\tilde\Omega = \sqrt{\Omega(\Omega + UN)}$ is one of the high Bogoliubov frequencies of the four-mode systen and
\begin{align}\label{J}
	\hat{J}\equiv\op[\rm L]{b}\o[\rm L]{b}+\op[\rm R]{b}\o[\rm R]{b}
\end{align}
the total joson number. In the terms of Hamiltonian adiabatic theory, $\hat{J}$ is an adiabatic invariant. Thus by maintaining our previous RWA we can now compute low frequency dynamics to leading order in small frequency ratios, dropping all terms in $\o{H}$ that do not commute with $\hat{J}$. The total atom numbers on each side, however, are not separately conserved for nonzero coupling $\omega$. Expressing the coupling term in terms of the Bogoliubov modes therefore forces us to again define new bosonic annihilation and creation operators $\o[\alpha]{A}$, $\op[\alpha]{A}$ that fulfill $\op[\alpha]{A} \o[\alpha]{A} = \o[\alpha]{N}$. These operators, which commute with the previously defined $\op[\alpha]{a}$, $\o[\alpha]{a}$ (and therefore naturally also with the Bogoliubov transformed quantities) enable us to take into account the rearrangement of atoms between the two subsystems. In these terms the coupling Hamiltonian $\o[\rm LR]{H}$ assumes the form
\begin{align}\label{HLR1}
\o[\rm LR]{H} = &-\frac{\omega}{2}\left(\sqrt{1-\frac{\op[\rm L]{a} \o[\rm L]{a}}{\o[\rm L]{N}}}\op[\rm L]{A} \o[\rm R]{A}\sqrt{1-\frac{\op[\rm R]{a} \o[\rm R]{a}}{\o[\rm R]{N}}}\right.\\
 &+  \left.\frac{1}{\sqrt{\o[\rm L]{N}}}\op[\rm L]{a} \o[\rm R]{a} \op[\rm L]{A} \o[\rm R]{A} \frac{1}{\sqrt{\o[\rm R]{N}}}+ \textrm{H. c.}\,\right),\nonumber
\end{align}
where by applying the same RWA as before we may write 
\begin{align}
\op[\alpha]{a} \o[\alpha]{a} = (u_\alpha^2 + v_\alpha^2)\op[\alpha]{b} \o[\alpha]{b},\quad
\op[\rm L]{a} \o[\rm R]{a} = u_{\rm L} u_{\rm R} \op[\rm L]{b} \o[\rm R]{b} + v_{\rm L} v_{\rm R} \op[\rm R]{b} \o[\rm L]{b}.\nonumber
\end{align}
Assuming that our system is in an eigenstate of the conserved total atom number $\o{N} = \o[\rm L]{N}+\o[\rm R]{N}$ with eigenvalue $N$, and of the total joson number $\hat J$ with eigenvalue $J$, we can now perform two final HPT. One to express $\o[\alpha]{N}$ in terms of  $N$ and an $N$-conserving operator $\o{a}$ that transfers atoms between the subsystems defined by
\begin{align}
 \o[\rm L,R]{N} &= \frac{1}{2}[ N \pm N^{1/2}(\op{a} + \o{a})] + \mathcal{O}(N^{-1/2})
\end{align}
and a completely analogous one to express $\o[\alpha]{b}$ in terms of $\hat{J}$ and a $J$-conserving operator $\o{b}$ that transfers josons between the subsystems defined by
\begin{align}
 \hat{J}_{\rm L,R} &= \frac{1}{2}[ J \pm J^{1/2}(\op{b} + \o{b})] + \mathcal{O}(J^{-1/2}).
\end{align}
To identify collective excitations, we finally linearize $\o{H}$ in $\o{a}$ and $\o{b}$ to obtain for fixed values of $N,J$
\begin{align}\label{Hfinal}
 \o[\textrm{lin}]{H} = \;&\omega\op{a}\o{a} + \frac{U}{4} N(\op{a} + \o{a})^2 + \omega_J \op{b}\o{b} + \frac{U_J}{4} J (\op{b} + \o{b})^2\nonumber\\
 &+ \frac{U}{2}\frac{\Omega}{\tilde\Omega}(NJ)^{1/2}(\op{a} + \o{a})(\op{b} + \o{b}),
\end{align}
with the definitions
\begin{align}\label{wJUJ}
\omega_{J}&\equiv \omega\frac{\Omega+UN/2}{\sqrt{\Omega (\Omega + UN)}}, \quad U_{J}&\equiv -\frac{U}{2}\frac{4\Omega +UN}{\Omega + UN}\;.
\end{align}
The first line of \eqref{Hfinal} shows two different forms of the standard single JJ Hamiltonian: one providing Josephson oscillations of atoms between the L and R subsystems, with Josephson frequency $\tilde\omega = \sqrt{\omega(\omega+UN)}$; and another one implying Josephson oscillations of josons, \ie second Josephson oscillations, with the frequency $ \tilde{\omega}_{J}=\sqrt{\omega_{J}(\omega_{J}+U_{J}J)}$. These two different types of Josephson modes are exactly analogous, except that, as mentioned before, $U_{J}$ and $U$ are of opposite sign. The second line of \eqref{Hfinal} finally couples the atom and joson modes, and arises because the frequency of fast Josephson oscillations in each subsystem depends on the respective atom number.

\section{Full quantum dynamics}

\begin{figure}[htb]
 \begin{center}
 \includegraphics[width=\linewidth]{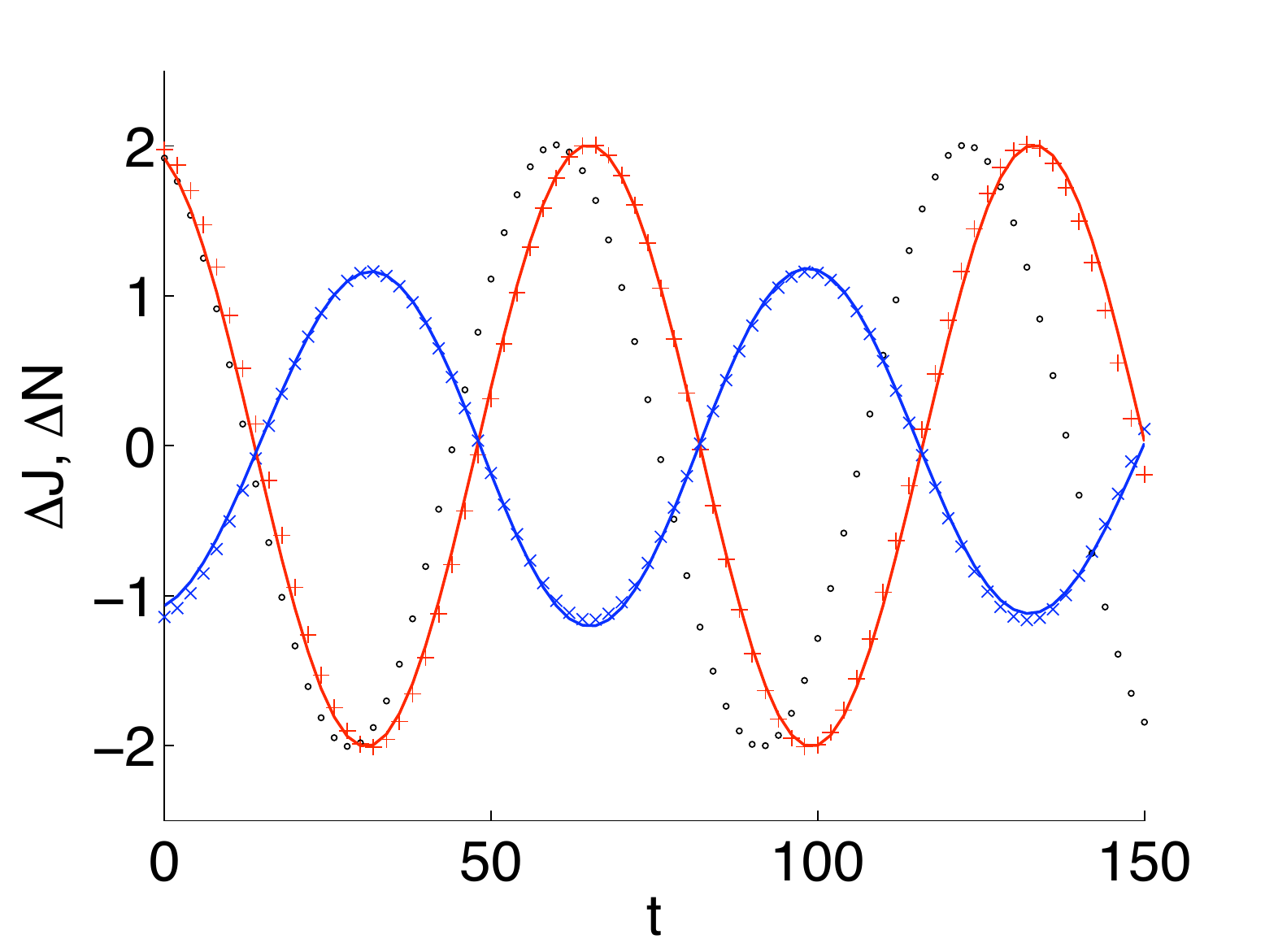}
 \end{center}
 \caption{Time evolution of $\Delta J =  \langle \op[\rm L]{b} \o[\rm L]{b} - \op[\rm R]{b} \o[\rm R]{b}\rangle$ and $\Delta N = \langle \o[\rm L]{N} - \o[\rm R]{N}\rangle$ of the full quantum system for the parameters $\Omega = 1$, $\omega = 0.1$, $UN = 0.1$, $N=44$ and $J = 5$.  Red $+$: numerical $\Delta J$. Red solid line: oscillation at second Josephson frequency $\tilde{\omega}_-$ computed in the text, fitted for phase.  Black $\circ$: oscillation at beat-frequency $\omega_J$. Blue $\times$: numerical $\Delta N$. Blue solid line: oscillation at a beat of  the frequencies $\tilde{\omega}_+$ and $\tilde{\omega}_-$ fitted for amplitude and phase.}
 \label{quantumnumerics}
\end{figure}

The crucial test of the joson theory will now be comparison to actual quantum dynamics, in particular in regard to the frequencies of collective excitations. The two modes appearing in the final Hamiltonian \eqref{Hfinal} derived in the previous section are still coupled, but they may easily be decoupled to yield the collective mode frequencies predicted by our analytical adiabatic theory:\cite{Strz10}
\begin{equation}\label{finalmodes}
 \tilde{\omega}_\pm^{2} = \frac{ \tilde{\omega}^2+ \tilde{\omega}_{J}^2}{2} \pm \left[\left(\frac{ \tilde{\omega}^2 -  \tilde{\omega}_J^2}{2}\right)^{2} +\frac{\omega\omega_J \Omega U^{2}NJ}{\Omega + UN}\right]^{1/2}\!\!\!\!\!\!\!\!.
\end{equation}
We will then test our adiabatic approximations by comparing this prediction with the results of numerical exact diagonalization of the full four-mode Hamiltonian. Since the dimension of the Hilbert space is only $\textrm{Dim}(\mathcal{H}) = \mathcal{O}(N^3)$, with destktop computers we may handle total particle numbers of more than $40$. Because $J$ is an adiabatic invariant, though not an exactly sharp quantum number, the expectation values of $\hat{J}$ in exact eigenstates of the Hamiltonian are all very close to integers, and possess very small quantum uncertainties. Therefore we may unambiguously assign $(N,J)$ labels to every eigenstate by setting $J$ to be the rounded value of $\langle\hat{J}\rangle$. To numerically check our joson theory we then take a superposition of the two lowest energy eigenstates for given values of $N$ and $J$ as initial state, and evolve it in time. Then as a function of time we plot the expectation value of the atom difference $\Delta N = \langle \o[\rm L]{N} - \o[\rm R]{N}\rangle$ and the joson difference $\Delta J = \langle \op[\rm L]{b} \o[\rm L]{b} - \op[\rm R]{b} \o[\rm R]{b}\rangle$. A typical example of the resulting evolution is shown,  or $N=44$ particles, in Fig.~\ref{quantumnumerics}.
We observe essentially perfect agreement of our analytically derived collective mode frequencies with the numerical results. This is not trivial, since what we have derived amounts to a free quasiparticle theory for excitations around higher excited states of our system, rather than around the ground state. The josons indeed oscillate between the subsystems with second Josephson frequency $\tilde \omega_-$, which is the lowest frequency in the system, and in particular is shifted well below the Bogoliubov beat frequency $\omega_J$ (shown in our plots for comparison), proving that the second Josephson oscillation is sensitive to post-Bogoliubov nonlinear dynamics. Since atoms and josons are coupled, our initial state also includes an excitation of the first Josephson oscillation of the atoms. 

\begin{figure}[htb]
 \begin{center}
 \includegraphics[width=\linewidth]{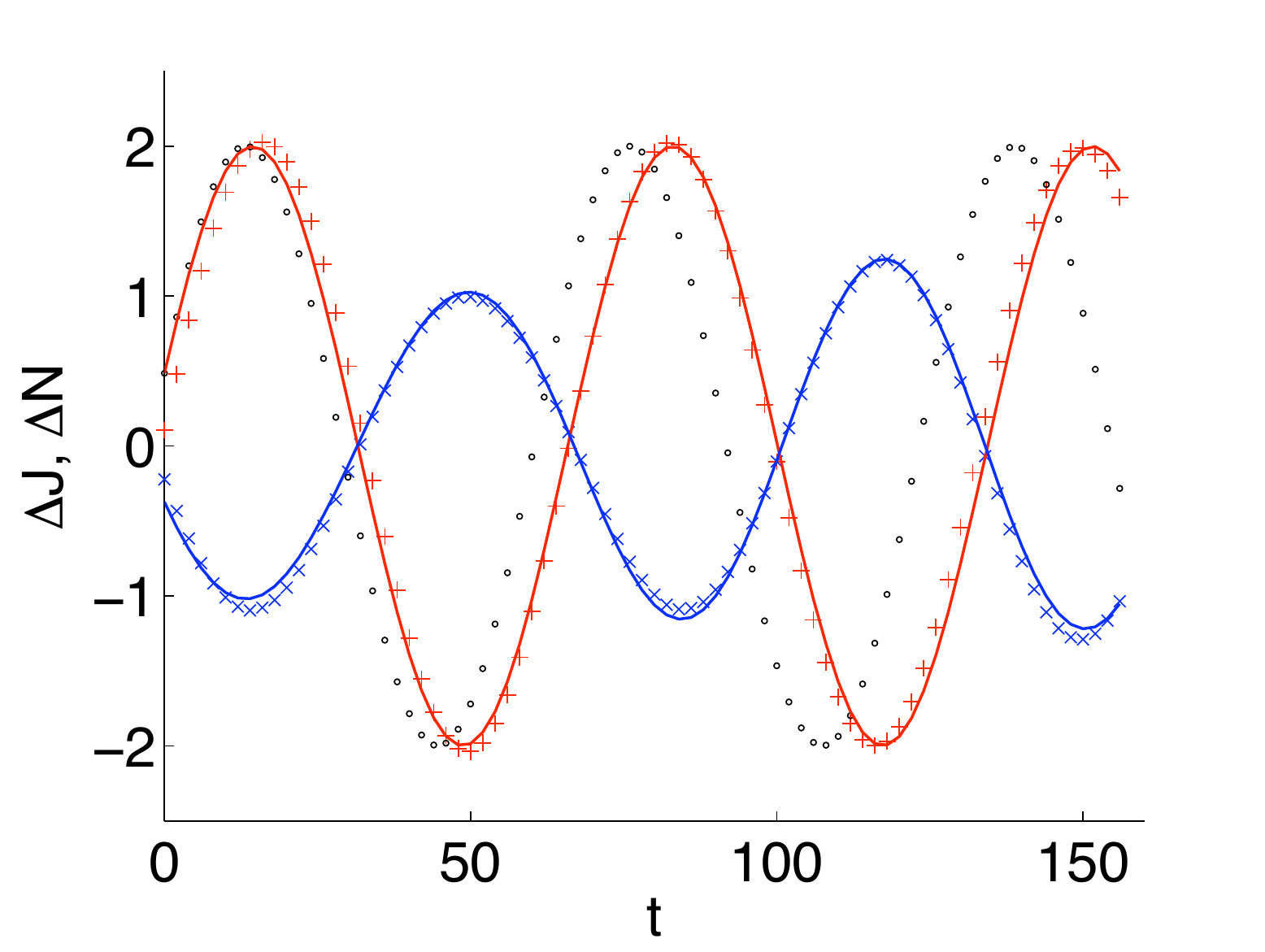}
 \end{center}
 \caption{Time evolution of $\Delta J$ and $\Delta N$ of the corresponding mean field system for exactly the same parameters as used in Fig.~\ref{quantumnumerics}. }
 \label{meannumerics}
\end{figure}

For another comparison we can also show the results of mean field evolution of our system, under a four-mode Gross-Pitaevskii nonlinear Schr\"odinger equation. In contrast to the dynamical procedure used in \cite{Strz10} to excite a mean field state with some value of $J$,  we now directly translate a given joson number to a certain particle imbalance, by inverting our series of Bogoliubov transformations, in order to construct an initial mean field state with an exact value of $J$. The resulting mean field dynamics for the same parameters as in the quantum case is shown in Fig.~\ref{meannumerics}.
Here again, the agreement with the linearized joson theory is excellent, as already shown in \cite{Strz10}. We may also confirm that for $N=44$ particles the agreement between mean field theory and the exact many particle dynamics is already very good. To explore instead the less classical regime of small $N$, while keeping large enough $J$ to have josons present to oscillate, the condition $J\ll N$ must be abandoned. This requires relaxing several simplifying assumptions made in the derivation in section \ref{section:joson}, in order to construct a more general theory of collective excitations including josons, which will be valid for small $N$ and for $J\sim N$ \cite{Strz12}.

\section{Four-mode quantum phase model}

The standard procedure of deriving a QPM for a single JJ starts with the two-mode Gross-Pitaevskii equation for the classical (mean field) variables, which is obtained from the quantum mechanical Heisenberg equation of motion by replacing all the operators with c-numbers. The procedure is to represent this classical theory in polar canonical variables, obtaining a Hamiltonian function with a cosine-potential. Re-quantizing canonically then leads to the Hamiltonian operator of the QPM, for which the corresponding Schr\"odinger equation is a Mathieu equation. Because canonical quantization in polar co-ordinates is not generally valid, the QPM Hamiltonian operator is an approximation, and is only accurate in the large $N$ limit. Unlike the mean field theory on which it was based, however, the QPM can express quantum superpositions, including the Gaussian squeezing that represents quantum depletion, as well as highly non-classical `running states'.

We therefore now construct a QPM for our four-mode system, which will respect the existence and conservation of josons as well as of atoms. We start with the Hamiltonian obtained above after the first HPT and Bogoliubov transformation, given by the sum of \eqref{Halpha1} and  \eqref{HLR1}. The corresponding mean field Hamiltonian function is obtained by replacing the operators according to 
\begin{align}
\o[\alpha]{A} \rightarrow \sqrt{R_{\alpha}}\e^{-\i\vartheta_{\alpha}},\quad \o[\alpha]{b} \rightarrow \sqrt{r_{\alpha}}\e^{-\i\beta_{\alpha}}
\end{align}
for $\alpha = {\rm L,R}$. In terms of these new canonically conjugate pairs $(R_\alpha,\vartheta_\alpha)$ and $(r_\alpha, \beta_\alpha)$ the Hamiltonian function becomes
\begin{align}
\mathcal{H} =& \sqrt{\Omega(\Omega+2UR_\alpha)}r_\alpha + \frac{U}{2}R_\alpha^2 - \frac{U}{2}\frac{2\Omega+UR_\alpha}{\Omega+2UR_\alpha}r_\alpha^2\nonumber \\
&-\omega\sqrt{r_{\rm L}r_{\rm R}}\Big[u_{\rm L}u_{\rm R}\cos(\vartheta_{\rm R}-\vartheta_{\rm L}+\beta_{\rm R}-\beta_{\rm L})\\
&+v_{\rm L}v_{\rm R}\cos(\vartheta_{\rm R}-\vartheta_{\rm L}-\beta_{\rm R}+\beta_{\rm L})\Big]\nonumber\\
&-\omega \left[(R_{\rm L}-(u_{\rm L}^2+v_{\rm L}^2)r_{\rm L})(R_{\rm R}-(u_{\rm R}^2+v_{\rm R}^2)r_{\rm R})\right]^{1/2}\nonumber\\
&\times \cos(\vartheta_{\rm R}-\vartheta_{\rm L}),\nonumber
\end{align}
omitting constant terms. As in the standard procedure for a single JJ we now introduce relative populations 
$R = (R_{\rm L}-R_{\rm R})/2 = \Delta N/2$ and $r = (r_{\rm L}-r_{\rm R})/2 =\Delta J /2$ as well as relative phases $\vartheta = \vartheta_{\rm L}-\vartheta_{\rm R}$ and $\beta = \beta_{\rm L}-\beta_{\rm R}$ of atoms and josons as new conjugate variables. Expanding to second order in $R$ and $r$, and assuming the high occupancy limit for atoms and josons, the Hamiltonian function may be cast into the form
\begin{align}\label{HQPM0}
\mathcal{H}_{\rm QPM} =&\; UR^2 + U_J r^2 +2U\frac{\Omega}{\tilde \Omega}rR -\frac{\omega}{2}(N-\tilde J)\cos\vartheta\nonumber\\
&-\frac{\omega_J}{2}J\cos\vartheta\cos\beta +\frac{\omega}{2}J\sin\vartheta\sin\beta
\end{align} 
where we have used the abbreviations $\omega_J$ and $U_J$ as introduced in \eqref{wJUJ}, and defined the new quantity $\tilde J \equiv (\Omega+UN/2)J/\tilde \Omega$. We then follow the standard procedure to re-quantize $\mathcal{H}_{\rm QPM}$ canonically. Since $\vartheta$, $R$ and $\beta$, $r$ are pairs of canonically conjugate variables, the QPM approach is to do this simply by replacing them with operators expressed as 
\begin{align}
\hat \vartheta = \vartheta, \quad \hat R = -\i \partial_{\vartheta}, \quad \hat \beta = \beta, \quad &\hat r = -\i \partial_{\beta}.
\end{align}
This leads to the Schr\"odinger equation $\i\partial_t \Psi = \o[\rm QPM]{H}\Psi$ for the wave function $\Psi(\vartheta,\beta)$. Since $\vartheta$ and $\beta$ are phases, $\Psi(\vartheta,\beta)$ must be $2\pi$-periodic in both arguments. In the framework of this QPM we are now able to analyze not only Josephson states, with the support of $\Psi$ localized in the potential wells around $\vartheta=\beta=0$, but also highly excited running states. 

To calculate the frequencies of excitations around the most localized Josephson states, we may simply expand the cosine-potentials up to second order and diagonalize the resulting coupled oscillator Hamiltonian. One of the two resulting frequencies $\sqrt{\lambda_\pm}$, however, is imaginary, since $\lambda_-<0$ while still $\lambda_+>0$. This may be understood easily, since explicitly performing the diagonalization shows that these frequencies indeed coincide with the quantum phase model limit $J,N\to\infty$ of the frequencies \eqref{finalmodes}, which can be written as $\tilde \omega_\pm^2 = \omega^2 + \lambda_\pm$ up to higher order corrections. We already know from the linearized theory reviewed above that because the inter-joson and inter-atom interactions are opposite in sign, either the first or second Josephson mode will become dynamically unstable for high enough $J$ or $N$. What linearization of the QPM around the phase-localized Josephson states has now shown is that system must already have reached this dynamical instability threshold when the QPM limit applies. The standard QPM can therefore never describe both forms of Josephson oscillation. 

We may nevertheless usefully apply the QPM to our system, to describe the non-phase-localized states that do become stable in the large $J,N$ regime. For simplicity we consider only the case of repulsive inter-atomic interactions, with attraction between josons. With $U_{J}<0$, the energy decreases with $\hat{r}^{2}$, so low energy favors the largest possible eigenvalues $\hat{r}^{2}$.  These eigenvalues can be as large as $J^{2}$, and so in the large $J$ limit where the QPM becomes accurate, the $U_{J}\hat{r}^{2}$ term in (\ref{HQPM0}) will always dominate the low-energy states. We therefore treat the coupling terms between josons and atoms in (\ref{HQPM0}) as a perturbation, and assume that to zeroth order the wave function $\Psi$ may be decomposed as the product of a running state for the josons and a phase-localized state for the atoms, of the form
\begin{align}
\Psi_{rn}(\vartheta,\beta) = \frac{1}{\sqrt{2\pi}}\e^{\i r\beta}\varphi_n(\vartheta).
\end{align}
Here $r$ is the eigenvalue of $\hat{r}$, and the low energy states for fixed $J$ will have $\vert r \vert \sim J \gg 1$. These $J$ `running' eigenstates of $\hat{r}$ represent states with many high-frequency joson excitations, in which the majority of josons remain permanently localized in either the L or R half of the four-mode system, depending on the sign of $r$. This corresponds to insulator-like quantum number locking in the second Josephson mode, while the first Josephson mode retains its usual phase-localized behavior.

In a Born-Oppenheimer sense we may consider $r$ to be sharp, and apply an $r$-dependent adiabatic Hamiltonian $\o[r]{H}$ for the first Josephson mode alone:
\begin{align}
\o[r]{H} &= \int_0^{2\pi} \Psi_{rn}^*(\vartheta,\beta) \o[\rm QPM]{H} \Psi_{rn}(\vartheta,\beta) \,\d \beta\\
&= U\hat\varrho^2 - \frac{\omega}{2}(N-\tilde J)\cos\vartheta+ r^2\left(U_J-U\frac{\Omega^2}{\tilde \Omega^2}\right)\nonumber
\end{align}
where we have defined the shifted atom imbalance $\hat\varrho \equiv -\i \partial_\vartheta + r\Omega/\tilde \Omega$. Thus the effective Schr\"odinger equation for the first Josephson mode, in a number-locked state of josons with given $r$, is again the Mathieu equation of the standard QPM, with the eigenstates given by Mathieu functions. There is a nontrivial effect from the josons on the atoms, however, in that the atom population imbalance $\Delta N = 2\langle \hat{R}\rangle$ is shifted away from zero by an amount proportional to the the joson imbalance $r$:
\begin{align}\label{shift}
\Delta N = 2\langle \hat \varrho \rangle - 2r\frac{\Omega}{\tilde \Omega} = 2\langle \hat \varrho \rangle - \Delta J  \frac{\Omega}{\tilde \Omega}.
\end{align}
This implies that if the mutually attractive josons are concentrated in the L subsystem, then the mutually repulsive atoms are partially driven toward the R subsystem; and vice versa.

From this we conclude that josons exert something like pressure on the atoms. This further supports our tentative, toy-model interpretation of josons as quanta of heat, since a hot gas should exhibit thermal pressure. Collecting josons on one side means heating this subsystem, according to our dynamical toy model for microscopic thermodynamics as an adiabatic theory.  It seems consistent with our picture, therefore, that the hotter gas should reveal its higher pressure, by forcing more atoms into the other subsystem. 

Of course, given that a certain number of high-frequency joson quanta are localized in one half of the system, it is obvious that the system can lower its energy by having fewer atoms present in the half of the system where they will have to participate in high-frequency excitations. Precisely because this is such a simple effect, it is not clear that it is entirely an artifact of our toy model. It may perhaps be a fair representation of the microphysics of thermal pressure in more general systems.

\section{Discussion}

We have argued that time scale separation is the essential dynamical element in hydrodynamic collective excitations of many-body systems. Such excitations are very common; they include ordinary sound waves in air. In macroscopic cases they may be described accurately by a low-frequency effective theory in terms of thermodynamical fields like pressure and temperature. The most natural relationship to postulate between these classical thermodynamical quantities, and the quantum Hamiltonian microphysics that underlies them, is that the thermodynamic quantities are approximate adiabatic invariants in the full system dynamics. They are not changed by rapid microscopic motion, but only by much slower macroscopic kinetics, via processes like convection or diffusion. The way in which the simple description of these adiabatic degrees of freedom emerges is certainly not simple, however. Even when the low-frequency effective hydrodynamics may be accurately linearized, as for small-amplitude sound waves, this linear theory will typically represent some strong and complicated nonlinear effects at the microscopic level, for example in the collisions between particles that maintain local equilibrium.

With our four-mode BH model we have explicitly realized this general scenario for relating thermo- and hydrodynamics to Hamiltonian microphysics, by constructing a simple quantum many-body Hamiltonian with a clear time scale hierarchy. Simple as it is, this model nonetheless has a quite complicated spectrum and exhibits quite nontrivial dynamics. We were able to show that there exists a highly excited collective mode that represents the exchange of high-frequency energy (josons) between the two subsystems, with the exchange proceeding as an oscillation at low frequency.

In this paper we extended the analysis of this nontrivial collective mode beyond our previous mean field limit, and we have done so in two ways. First we extended our numerical studies to the quantum case, and showed excellent agreement of our linearized joson theory with the actual quantum evolution, for moderately large numbers.  We thereby confirmed that our second Josephson frequency directly manifests itself in the quantum spectrum of the complete four-mode system, as an excitation energy above the lowest energy eigenstate with fixed $N$ and $J$.  To extend our adiabatic analysis further into the quantum regime than this, it is necessary to relax the condition of large $N$ and $J\ll N$ that we assumed introducing the joson theory in \cite{Strz10} and also used here. The extended theory, valid for small $N$ and $J~N$, is described elsewhere \cite{Strz12}.

Secondly we applied the quantum phase model to our four-mode system, and by examining number-locked joson states represented by running states of the quantum phase model we were able to show that localized josons exert a kind of pressure on atoms, in keeping with their postulated role as quanta of heat. Of course, the fact that josons concentrate spontaneously, in the case of repulsive interactions among atoms, does not represent the usual diffusive behavior of heat. The attractive interactions of josons need not entirely rule out their heat interpretation, however, since it is well known that special kinds of dynamics can produce systems with negative specific heat capacity, \eg self-gravitating systems, for which heat does indeed spontaneously concentrate, without any violation of thermodynamic laws. 

For further tests of dynamical conjectures about thermodynamics, it will clearly be necessary to investigate larger systems with more degrees of freedom. In larger Bose-Hubbard systems there are, on the one hand, many new possibilities for the dynamics of josons. On the other hand, however, they in general tend to be dynamically chaotic, which might play an important role in the emergence of thermodynamics. If at some point the complexity of these models becomes theoretically completely intractable, it may still prove possible to investigate them by means of experiments on cold, trapped Bose gases. While such experiments would be difficult, they are clearly feasible in principle as an extension of demonstrated capabilities. They would be a fine example of the principle that we do not study physics to learn about cold atoms, but rather study cold atoms to learn about physics.


\end{document}